\documentclass[review]{elsarticle}   

\usepackage[T1]{fontenc}
\usepackage[utf8]{inputenc}
\usepackage{hyperref}
\usepackage{url}
\usepackage{booktabs}
\usepackage{amsfonts}
\usepackage{amsmath}
\usepackage{amssymb}
\usepackage{nicefrac}
\usepackage{microtype}
\usepackage{graphicx}
\usepackage{bm}
\usepackage{placeins}
\usepackage[table]{xcolor}
\definecolor{ourcol}{HTML}{FFE5E5}
\definecolor{bestcol}{HTML}{FFF2A8}
\definecolor{ourtext}{HTML}{7A0000}

\journal{Journal of Information Security and Applications}

\begin{document}

\begin{frontmatter}

\title{PassGPT+: Leveraging Linguistic Priors for Password Modeling}

\author[cbe]{Rajneesh Anand}
\author[sutd]{Neeraj Lakshmanan}
\author[cse]{Masoud Yari\corref{cor1}}
\ead{may322@lehigh.edu}
\cortext[cor1]{Corresponding author}
\affiliation[cbe]{organization={Department of Chemical and Biomolecular Engineering, Lehigh University},
  city={Bethlehem}, state={PA}, postcode={18015}, country={USA}}
\affiliation[sutd]{organization={Department of Computer Science \& Design, Singapore University of Technology \& Design},
  country={Singapore}}
\affiliation[cse]{organization={Department of Computer Science \& Engineering, Lehigh University},
  city={Bethlehem}, state={PA}, postcode={18015}, country={USA}}

\begin{abstract}
Passwords remain the dominant online authentication mechanism, and understanding how humans choose them is essential for defensive strength estimation and attack simulation alike.
Recent learning-based approaches such as PassGAN and PassGPT have shown that deep generative models can learn password structure directly from leaked corpora. However, both train from random initialization on password data alone.
The role of linguistic prior knowledge in password modeling, and what it reveals about how humans create secrets, remains largely underexplored.
Here, we address this gap with PassGPT+, which adapts the linguistic prior of GPT-2 to password observations through character-aware tokenization. We also  introduce PassDiffusion, the first absorbing-state discrete diffusion model for password generation, as a probe of whether non-autoregressive approaches are competitive. On the RockYou benchmark, PassGPT+ recovers $22.53\%$ of held-out passwords at $10^8$ guesses, a $16\%$ relative gain over PassGPT, and retains $79\%$ of this match rate when transferred without retraining to a disjoint 2020 leak dataset, demonstrating that linguistic priors capture persistent regularities of human password generation.
PassDiffusion underperforms by two to three orders of magnitude, indicating that autoregressive modeling is substantially better matched than iterative denoising to the discrete, exact-match nature of password generation.
\end{abstract}


\end{frontmatter}

\section{Introduction}

Passwords continue to serve as the most widely used authentication mechanism across digital platforms, despite the availability of alternative technologies such as biometrics and hardware tokens~\citep{wayman2005biometric, hunt2018passwords}. Their dominance is largely due to their simplicity, ease of deployment, and familiarity among both users and developers. However, this widespread reliance on passwords comes with a serious downside: large-scale password leaks have repeatedly shown that users tend to pick weak and predictable passwords~\citep{dell2010password, wang2017zipf}. As these leaked datasets grow in size and frequency, they provide attackers with a rich source of information that can be used to crack password hashes and compromise user accounts.

Traditionally, password guessing has been carried out using rule-based tools such as HashCat~\citep{hashcat} and John the Ripper~\citep{jtr}, which apply heuristics like dictionary attacks, leet speak transformations, and word concatenation to expand a base dictionary into a large pool of candidate passwords. While these tools are remarkably efficient at generating matches quickly, they are fundamentally limited by the scope of their hand-crafted rules. Extending them requires specialized expertise and manual effort, and each rule set can only capture a specific subset of the password space~\citep{weir2009password, durmuth2015omen}.

To move beyond these limitations, researchers have turned to deep learning. \citep{hitaj2019passgan} introduced PassGAN, the first approach to use Generative Adversarial Networks (GANs) for password guessing. PassGAN uses the Improved Wasserstein GAN (IWGAN)~\citep{gulrajani2017improved} framework to learn the distribution of real passwords from the RockYou leak without requiring any prior knowledge about password structures. Although PassGAN showed that machine learning could autonomously discover password patterns, it suffered from issues such as low sample uniqueness and the inability to provide explicit probability estimates over generated passwords.

More recently, \citep{rando2023passgpt} proposed PassGPT, an autoregressive language model based on the GPT-2 architecture~\citep{radford2019language} that is trained from scratch on password leaks. PassGPT addressed several of PassGAN's shortcomings by modeling the conditional distribution of characters in a password, enabling features like guided generation under arbitrary constraints and explicit probability estimation. PassGPT was shown to outperform PassGAN by guessing roughly $20\%$ more previously unseen passwords. Follow-up work such as PagPassGPT~\citep{su2024pagpassgpt} further improved upon PassGPT by incorporating pattern structure information to guide the generation process. Other approaches have explored normalizing flows~\citep{pagnotta2022passflow}, recurrent neural networks~\citep{melicher2016fast}, and more recently, adversarial ranking frameworks~\citep{yang2025rankguess} for password guessing.

However, a key aspect that remains underexplored in the existing literature is the role of prior beliefs in password modeling. Both PassGAN and PassGPT train their respective architectures from random initialization using only password data. This means that models must learn all sequential pattern recognition from scratch, ignoring the rich linguistic structures and semantic rules that humans naturally rely on when creating passwords. Without incorporating this prior knowledge into the statistical model, existing models may struggle to generalize beyond the specific datasets on which they were trained, missing the underlying cognitive logic that often drives password formulations~\citep{radford2018improving, devlin2019bert, brown2020language}.

\paragraph{Our contributions.} In this work, we investigate whether the priors encoded in the foundation models extend to the domain of password guessing. We show that considering the  prior knowledge encoded in the foundation model on password observations in training our model significantly outperforms PassGPT's from scratch strategy, demonstrating that the sequential structure encoded in large-scale language priors captures regularities of human-chosen passwords. 
We provide a systematic comparison across the seven approaches: PassGAN, PassGAN*, PassGPT, PassVQT, HashCat Best64, PassDiffusion, and our model PassGPT+ on the RockYou benchmark, analyzing match rates, uniqueness of generated passwords, and the role of training configuration choices such as the number of update epochs and the inclusion of duplicate passwords. We discuss the architectural and inductive-bias differences that drive the observed performance gains, offering practical guidance for future work on foundation-model-based password modeling.

\begin{figure}[htbp]
    \centering
    \includegraphics[width=\linewidth]{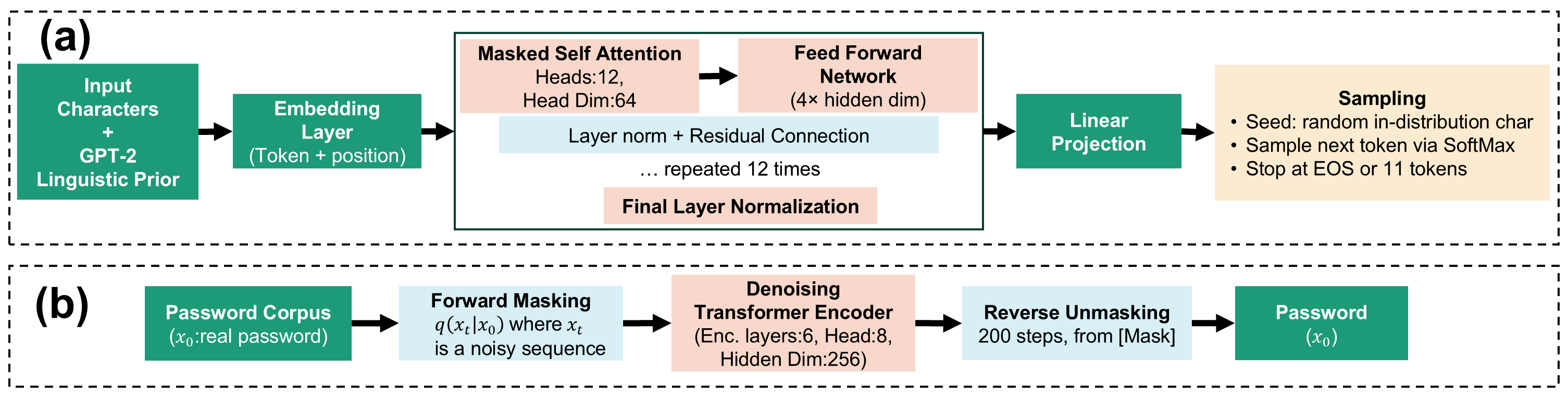}
    \caption{Proposed architectures for password generation. \textbf{(a)} PassGPT+: a 12-layer causal Transformer decoder that adapts a variant of GPT-2 architecture and is trained using its weights through character-level tokenization. Input characters are mapped to GPT-2 BPE token IDs and processed through 12 masked self-attention blocks (12 heads, head dim 64) with feed-forward expansion ($4\times$ hidden dim), layer normalization, and residual connections. Generation is seeded with a random in-distribution character rather than the standard BOS token, avoiding out-of-distribution initialization. \textbf{(b)} PassDiffusion (or PassDiff): an absorbing-state discrete diffusion model for password generation. A real password $x_0$ is corrupted via forward masking $q(x_t \mid x_0)$ over $T=1000$ timesteps, producing a progressively masked sequence $x_t$. A bidirectional Transformer encoder (6 layers, 8 heads, hidden dim 256) conditioned on timestep $t$ learns to predict $x_0$ from $x_t$. At inference, passwords are generated by iterative reverse unmasking over 200 steps from a fully masked sequence.}
    \label{fig:fig1}
\end{figure}

\section{Experimental setup}
\label{sec:experimental_setup}

In this section, we describe the datasets, model architectures, training procedures, and evaluation metrics used in our study across seven different password generation approaches: a rule-based baseline (Hashcat), two GAN-based models (the original PassGAN and the representation-learning-enhanced PassGAN*), a vector-quantized transformer baseline (PassVQT), our new framework PassDiffusion (a diffusion model for passwords), and two autoregressive transformer models: the original PassGPT trained from scratch and our model, which we call here PassGPT+.

\subsection{Datasets}
\label{sec:datasets}

\paragraph{RockYou dataset (primary benchmark)}
We use the RockYou password corpus~\citep{burns_rockyou} as our primary benchmark, a standard dataset in password generation research~\citep{hitaj2019passgan, rando2023passgpt, su2024pagpassgpt}. Following the established protocol from both PassGAN~\citep{hitaj2019passgan} and PassGPT~\citep{rando2023passgpt}, we filter the dataset to retain only passwords of at most 10 characters. After filtering, the dataset contains approximately 11.9 million password entries. We split this filtered corpus into training and test sets using an $80/20$ ratio. Before splitting, the entire dataset is shuffled to eliminate any ordering bias. The test set is then cleaned by removing any password that also appears in the training split, ensuring strict non-overlap between the two sets. This procedure yields $9{,}519{,}331$ training passwords and $2{,}381{,}822$ unique test passwords. All models are evaluated against this held-out test set, ensuring a fair comparison.

\paragraph{2020 leaked password dataset (cross-distribution evaluation)}
The RockYou corpus reflects password habits from 2009, and it is reasonable to ask whether models trained on it can generalize to more modern password distributions. To investigate this, we construct a second evaluation set from a large-scale 2020 credential leak \citep{ignis_pwdb} comprising approximately 10 million plaintext passwords. To make sure this dataset contributes genuinely new test signal and does not overlap with our training data, we first remove all passwords that appear anywhere in the RockYou corpus, eliminating any possibility of cross-dataset leakage. We then apply the same length filter ($\leq 10$ characters). From the resulting filtered corpus, we randomly sample $2{,}381{,}844$ passwords, which matches the size of the RockYou test set to allow direct comparability between the two evaluation benchmarks. The PassGPT+ model is also evaluated against this 2020 dataset without any retraining or fine-tuning. This provides a cross-distribution generalization benchmark that reflects modern password-creation patterns and lets us assess how well the PassGPT+ model captures universal password structures versus RockYou-specific idiosyncrasies.

\subsection{Model architectures}
\label{sec:models}

We develop three password generation approaches and compare it against another four password generation approaches, ranging from a traditional rule-based tool to deep generative models.

\subsubsection{Hashcat (rule-based baseline)}
\label{sec:hashcat}

As a non-learning-based reference point, we use Hashcat~\citep{hashcat}, which is the industry-standard password recovery tool widely used by both security researchers and penetration testers. Hashcat generates candidate passwords by applying hand-crafted transformation rules such as dictionary attacks, leet speak substitutions (e.g., \texttt{password} $\rightarrow$ \texttt{p4s5w0rd}), word concatenation, and case toggling to a base dictionary~\citep{weir2009password}. For our evaluation, we generate passwords using Hashcat's built-in rule-based attack modes and sample unique passwords from the output. These are then evaluated against both test sets under the same match-rate metric as all other models. Hashcat serves as a practical baseline representing the current state of the art in rule-based password cracking.

\subsubsection{PassGPT+}
\label{sec:passgptplus}

 Our primary model architecture is shown in Figure~\ref{fig:fig1}(a), which we call \textbf{PassGPT+}, which fine-tunes the publicly available pre-trained GPT-2 model~\citep{radford2019language} on the password training set. GPT-2 is a 117-million-parameter decoder-only transformer that was originally pre-trained on approximately 40 GB of web text. The key idea behind our approach is that, unlike the original PassGPT~\citep{rando2023passgpt} which trains a GPT-2-style architecture from random initialization using only password data, we initialize from pre-trained GPT-2 weights and fine-tune them on passwords. This leverages the rich sequential pattern knowledge that GPT-2 already learned from natural language such as character co-occurrence patterns, common n-grams, and general sequence structure which, as we will show, capture much of the structure of human-chosen passwords.

\paragraph{Tokenization.} One of the important design decisions in our model is how we handle tokenization. GPT-2's default tokenizer uses byte-pair encoding (BPE)~\citep{radford2019language}, which is a subword tokenization scheme designed for natural language. If we were to use BPE directly on passwords, it would segment them unpredictably; for example, the password \texttt{monkey123} might be tokenized as \texttt{[mon, key, 123]} or \texttt{[monkey, 12, 3]}, depending on what subwords the tokenizer learned from its web text training data. This is undesirable for password modeling, where we want character-level control. To solve this, we adopt a character-level tokenization scheme that works within the existing GPT-2 vocabulary. We build a simple lookup table that maps each unique ASCII character in the training set directly to a single GPT-2 BPE token ID. For instance, the character \texttt{`a'} might map to token ID 64, the character \texttt{`1'} to token ID 16, and so on. Each password is then encoded as a sequence of these per-character token IDs, followed by an end-of-sequence (EOS) token (ID 50256). All sequences are padded to a fixed length of 12 tokens (10 characters + 1 EOS + 1 buffer). The padding positions are assigned a label of $-100$ in the training targets, which tells PyTorch's cross-entropy loss function to ignore these positions. This way, the model only learns to predict actual password characters and the EOS token.

\paragraph{Training configuration.} The model is fine-tuned for 3 epochs using the AdamW optimizer~\citep{loshchilov2019decoupled} with a learning rate of $5\times 10^{-5}$ and a batch size of $512$. We use mixed-precision (FP16) training to reduce GPU memory and accelerate throughput. The training is conducted on the full $\sim$9.5M-password RockYou \citep{burns_rockyou} training set (with duplicates), and the total training time is 30 hours on a NVIDIA RTX 5060 (8GB VRAM) GPU.

\paragraph{Generation procedure.} In standard GPT-2 text generation, one would seed the model with a beginning-of-sequence (BOS) token (ID 50256). However, during our fine-tuning, the training sequences do not contain any BOS prefix; they start directly with the first password character. This means that if we seed generation with BOS, the model is placed in an out-of-distribution state and tends to produce subword patterns instead of passwords. To fix this, we seed each password generation with a single randomly sampled character drawn from the set $\{$\texttt{a-z, A-Z, 0-9, !@\#\_.}$\}$, characters that the model saw at the start of training sequences. Generation then proceeds autoregressively with temperature $\tau=1.0$ and top-$k=50$ sampling, stopping when either the EOS token is sampled or the sequence reaches 11 tokens.

\paragraph{Comparison with PassGPT.} Table~\ref{tab:passgpt_comparison} summarizes the architectural and training differences between the original PassGPT~\citep{rando2023passgpt} and our PassGPT+. The most important differences are: (i) we initialize from pre-trained GPT-2 weights rather than an uninformative prior over character sequences; (ii) we use the full 12-layer, 117M-parameter GPT-2 instead of PassGPT's smaller 8-layer variant; and (iii) we adapt to password observations over 3 update epochs instead of just 1.

\begin{table}[t]
\caption{Comparison between the model architecture of PassGPT~\citep{rando2023passgpt} and our PassGPT+. The three changes most relevant to performance are linguistic prior initialization, increased depth, and additional fine-tuning epochs.}
\label{tab:passgpt_comparison}
\centering
\small
\begin{tabular}{lll}
\toprule
\textbf{Aspect} & \textbf{PassGPT}~\citep{rando2023passgpt} & \textbf{PassGPT+ (Ours)} \\
\midrule
Initialization     & Uninformative (random init)        & GPT-2 linguistic prior \\
Tokenizer          & Custom 256-char         & GPT-2 BPE (char-mapped) \\
Layers             & 8 layers, 12 heads      & 12 layers, 12 heads \\
Parameters         & ${\sim}85$M             & 117M \\
Training epochs    & 1                       & 3 \\
Generation seed    & BOS (\texttt{<s>})      & Random password char \\
\bottomrule
\end{tabular}
\end{table}

\subsubsection{PassDiffusion (discrete diffusion model)}
\label{sec:passdiffusion}

We implement \textbf{PassDiffusion} (or, PassDiff), shown in Figure~\ref{fig:fig1}(b), a character-level password generator using Discrete Denoising Diffusion Probabilistic Models (D3PM) with absorbing states~\citep{austin2021structured}. To our knowledge, this is the first application of diffusion models to password generation. We emphasize that PassDiffusion is not a Diffusion Transformer (DiT)~\citep{peebles2023scalable}, which is a continuous-space image generation architecture. Our model operates entirely in discrete token space: the diffusion framework handles the corruption and denoising schedule, while a Transformer encoder serves as the denoising backbone.

\paragraph{Forward process.} Given a clean password $\bm{x}_0$ of up to 10 characters, the forward process independently replaces each token with a \texttt{[MASK]} token. The survival probability at timestep $t$ is $\bar{\alpha}_t = \prod_{s=1}^{t}(1-\beta_s)$, with a linear schedule ($\beta_1 = 10^{-4}$, $\beta_T = 0.02$, $T = 1000$). At $t=0$ the password is intact; at $t=T$ it is fully masked.

\paragraph{Denoising backbone.} The denoising network $f_\theta$ is a bidirectional Transformer encoder~\citep{vaswani2017attention}; unlike PassGPT+'s causal GPT-2 decoder, which generates left-to-right, our encoder attends to all token positions simultaneously. This bidirectionality is a natural fit for diffusion, where the model must reason about all masked positions jointly rather than sequentially. The architecture has 6 layers, 8 attention heads, hidden dimension 256, and pre-layer normalization, yielding 5.3M parameters trained from scratch. Timestep $t$ is encoded via a sinusoidal embedding projected through a two-layer MLP and broadcast-added to all token positions. The vocabulary contains 95 ASCII characters plus \texttt{[PAD]} and \texttt{[MASK]} (98 tokens total). Training minimizes cross-entropy loss on masked positions only~\citep{devlin2019bert}, using AdamW~\citep{loshchilov2019decoupled} ($\text{lr}=10^{-4}$, cosine decay, 10 epochs, batch size 512).

\paragraph{Generation.} Starting from a fully masked sequence $(\texttt{[MASK]})^{10}$, we apply the reverse process over 200 evenly-spaced steps. At each step $t \rightarrow t_{\text{next}}$, the model predicts $\bm{x}_0$, and each still-masked position is unmasked with probability $(\bar{\alpha}_{t_{\text{next}}} - \bar{\alpha}_t)/(1 - \bar{\alpha}_t)$. Once revealed, a character stays fixed. At $t=0$, remaining masks are resolved via argmax. The \texttt{[MASK]} and \texttt{[PAD]} tokens are suppressed at every step to ensure only valid characters appear.

\subsection{Evaluation metric}
\label{sec:eval_metric}

All models in our study are evaluated using the \textbf{match rate} metric, which is defined as
\begin{equation}
\text{Match Rate} = \frac{|\mathcal{G} \cap \mathcal{T}|}{|\mathcal{T}|} \times 100\%,
\label{eq:match_rate}
\end{equation}
where $\mathcal{G}$ is the set of unique passwords generated by a model and $\mathcal{T}$ is the set of unique passwords in the test set. This metric, adopted from PassGAN~\citep{hitaj2019passgan} and PassGPT~\citep{rando2023passgpt}, measures what fraction of real human-chosen passwords a given model can recover, which directly quantifies how effective the model would be in practical password cracking.

We report match rates at multiple generation sizes: $10^4$, $10^5$, $10^6$, $10^7$, and $10^8$ generated passwords. Evaluating at multiple generation sizes is important because it reveals how each model's performance scales with computational effort. A model that achieves high match rates at smaller generation sizes is more efficient and practically useful, while a model that keeps improving at larger generation sizes demonstrates a richer and more diverse coverage of the password space.

\section{Results}
\label{sec:results}

\subsection{Comparison with existing password guessing approaches}

Table~\ref{tab:comparison_main} compares the password match rate of the considered methods across different generation sizes. The compared baselines include rule-based guessing using Hashcat~\citep{hashcat}, GAN-based password generation using PassGAN~\citep{hitaj2019passgan}, transformer-based password modeling using PassGPT~\citep{rando2023passgpt}, and other reported methods such as PassVQT~\citep{rando2023passgpt}. Our main contributions are PassGPT+, an autoregressive model that adapts the linguistic prior of GPT-2 to password observations, and PassDiffusion, a discrete diffusion-based model.

At very small generation sizes ($10^4$--$10^5$ guesses), the differences among several learning-based methods are relatively small, and Hashcat remains highly competitive. This is expected because rule-based systems can efficiently capture the most common and easiest password patterns in the early-guess regime. For example, at $10^5$ guesses, Hashcat achieves a match rate of $0.0918\%$, which is slightly higher than PassGPT+ ($0.0847\%$). However, this trend changes as the generation size increases.

From $10^6$ guesses onward, PassGPT+ becomes the strongest overall method among the compared approaches. At $10^6$ guesses, PassGPT+ achieves a match rate of $0.78\%$, outperforming PassGPT ($0.50\%$), PassVQT ($0.45\%$), PassGAN ($0.38\%$), and Hashcat ($0.69\%$). At $10^7$ guesses, PassGPT+ reaches $5.66\%$, exceeding PassGPT ($4.25\%$) by about $33.1\%$ relative improvement and outperforming Hashcat ($4.65\%$) by about $21.6\%$. At the largest reported generation size of $10^8$ guesses, PassGPT+ achieves the best overall result of $22.53\%$, compared to $19.37\%$ for PassGPT, $10.30\%$ for PassVQT, $9.51\%$ for PassGAN*, and $6.73\%$ for PassGAN. These results show that adapting an autoregressive foundation model with linguistic priors leads to consistent gains, especially in the deeper-guess regime where modeling long-range sequential structure becomes more important.

\begin{table}[htbp]
\caption{Match rate (\%) on the RockYou test set across generation sizes. Best result per row is highlighted in yellow. Em-dashes (---) denote values not reported in the original publications. PassGPT+ wins at every generation size of $10^6$ or larger.}
\label{tab:comparison_main}
\centering
\small
\setlength{\tabcolsep}{4pt}
\resizebox{\linewidth}{!}{%
\begin{tabular}{lccccccc}
\toprule
\textbf{Guess} & \textbf{PassGAN} & \textbf{PassGAN*} & \textbf{PassVQT} & \textbf{PassGPT} & \textbf{PassDiffusion} & \textbf{HashCat} & \textbf{PassGPT+} \\
 & \citep{hitaj2019passgan} & \citep{pasquini2021improving} & \citep{rando2023passgpt} & \citep{rando2023passgpt} & (Ours) & & (Ours) \\
\midrule
$10^4$ & \cellcolor{bestcol}.010 & ---  & .0040 & .010  & .0000 & .0095 & {\color{ourtext}\textbf{.0093}} \\
$10^5$ & .050  & ---  & .050  & .050  & .0002 & \cellcolor{bestcol}.092 & {\color{ourtext}\textbf{.085}} \\
$10^6$ & .380  & ---  & .450  & .500  & .0018 & .690  & \cellcolor{bestcol}{\color{ourtext}\textbf{.780}} \\
$10^7$ & 2.04  & ---  & 2.90  & 4.25  & .030  & 4.65  & \cellcolor{bestcol}{\color{ourtext}\textbf{5.66}} \\
$10^8$ & 6.73  & 9.51 & 10.30 & 19.37 & ---   & ---   & \cellcolor{bestcol}{\color{ourtext}\textbf{22.53}} \\
\bottomrule
\end{tabular}%
}
\end{table}

\begin{figure}[htbp]
    \centering
    \includegraphics[width=\linewidth]{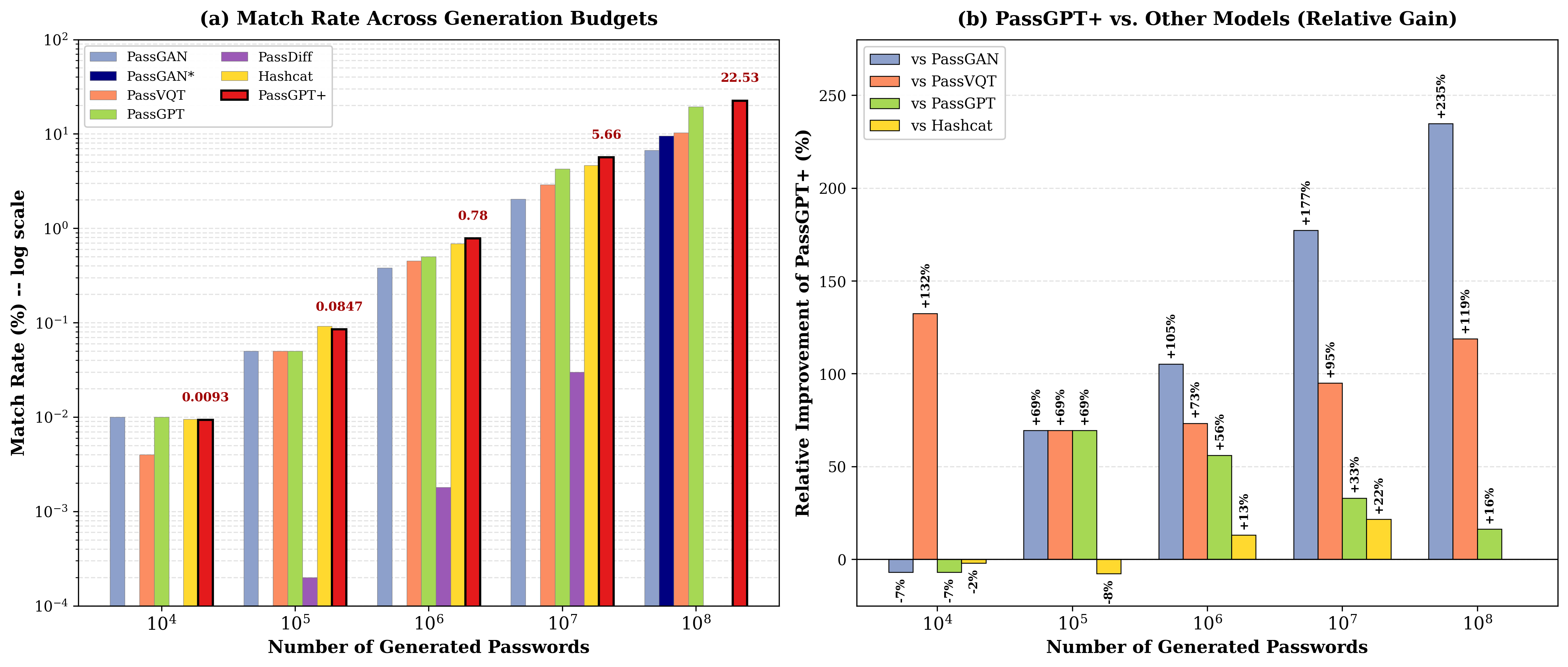}
    \caption{Comparison of PassGPT+ against all other models on the RockYou test set. \textbf{(a)} Match rate of each model on a logarithmic vertical axis. PassGPT+ (red, outlined in black) achieves the highest match rate at every generation size from $10^6$ guesses onward. The very low bars for PassDiffusion confirm that the discrete diffusion framework is poorly matched to exact-character password generation. \textbf{(b)} Relative improvement of PassGPT+ over each learning-based baseline and Hashcat, computed as $[(\text{PassGPT+} - \text{other model})/\text{other model}] \times 100\%$. Positive values indicate that PassGPT+ recovers more passwords than the baseline at that generation size. The gain over PassGPT (the closest competitor) grows from $+56\%$ at $10^6$ to $+16\%$ at $10^8$, while the gain over PassGAN exceeds $+200\%$ at the largest generation size. Hashcat marginally wins at very small generation sizes ($10^4$ and $10^5$), reflecting the well-known efficiency of hand-crafted rules at low guess counts, but PassGPT+ overtakes it from $10^6$ onward.}
    \label{fig:rockyou_comparison}
\end{figure}

A key observation from Figure~\ref{fig:rockyou_comparison} is that the performance improvement of PassGPT+ over PassGPT becomes especially meaningful once the number of guesses is sufficiently large. The relative gain is approximately $56.0\%$ at $10^6$ guesses, $33.1\%$ at $10^7$ guesses, and $16.3\%$ at $10^8$ guesses. This suggests that the proposed fine-tuning strategy does not only help recover the most frequent passwords, but also improves the quality of the longer candidate list. In practical password auditing scenarios, this is important because strong methods should remain effective not only in the first few guesses, but also when the generation size is extended.

\subsection{Performance of PassDiffusion}

In contrast to PassGPT+, the proposed PassDiffusion model performs poorly across all tested generation sizes (shown in Table~\ref{tab:comparison_main}). Its match rate is only $0.0002\%$ at $10^5$ guesses, $0.0018\%$ at $10^6$ guesses, and $0.03\%$ at $10^7$ guesses, which is two to three orders of magnitude lower than PassGPT+ over the same range. Although this result is negative from a performance perspective, it is still scientifically important because it highlights a mismatch between discrete diffusion mechanisms and the password generation problem.

A likely reason is that password guessing is an exact discrete prediction task: a password is either correct or incorrect, and even a one-character error makes the entire guess redundant. In diffusion-based generation, errors can accumulate during the repeated denoising process, especially when many characters are masked at high-noise timesteps. In addition, unlike foundation-model-based approaches such as PassGPT+, the diffusion model in our setting learns from an uninformative prior over character sequences and therefore does not benefit from any prior knowledge of sequential structure. Therefore, our results suggest that adapting an autoregressive linguistic prior is significantly better suited than discrete diffusion for password generation.

\subsection{Robustness on a newer leaked dataset}

We define retention at a given generation size as the ratio of the 2020 match rate to the RockYou match rate at that same generation size; this normalization isolates how much of the model's learned password-cracking capability transfers across distributions, independent of the absolute difficulty of either dataset. We mark a $50\%$ baseline because it represents the natural threshold at which a model retains at least half of its original strength on unseen data.

\begin{figure}[htbp]
    \centering
    \includegraphics[width=0.85\linewidth]{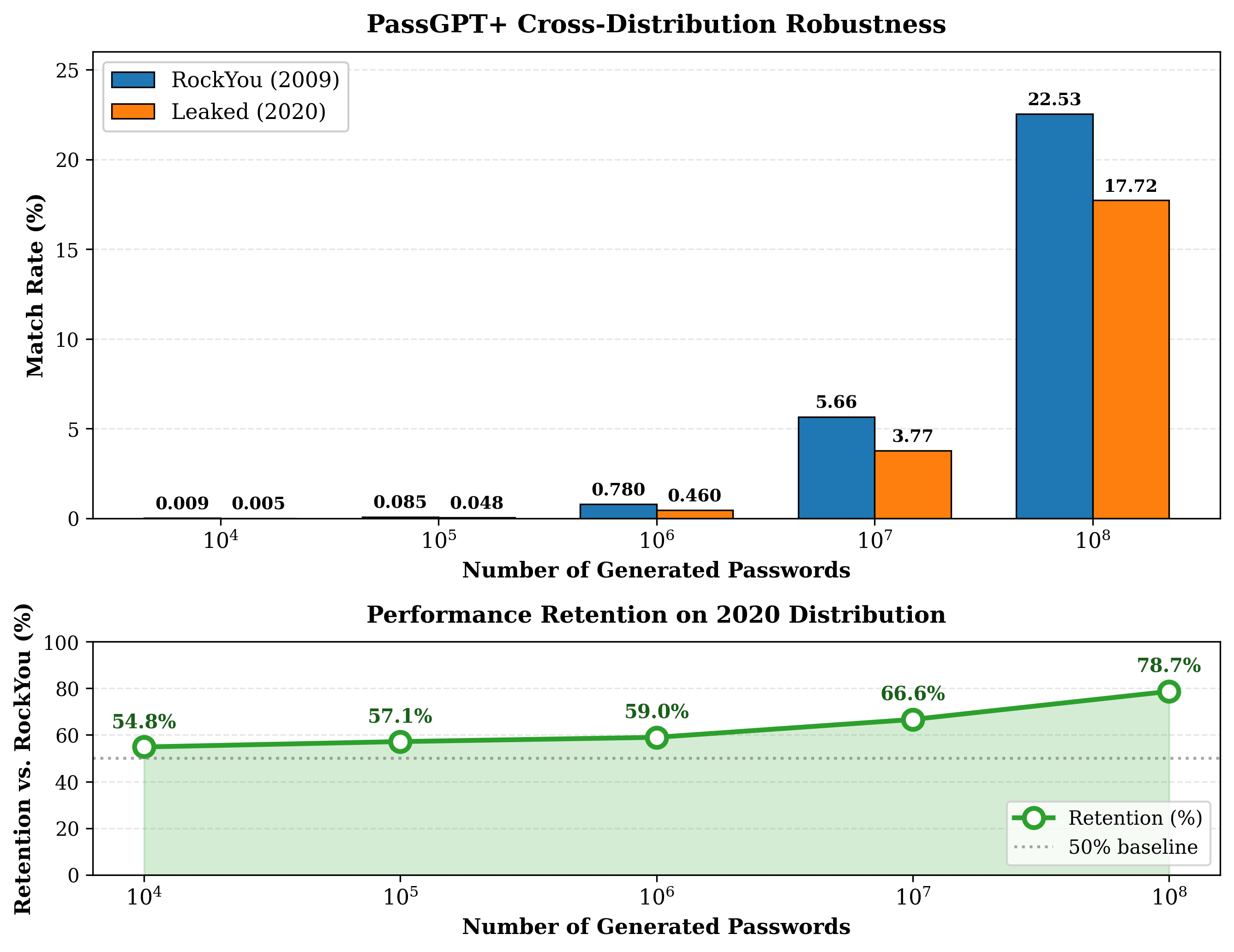}
    \caption{Cross-distribution robustness of PassGPT+ evaluated on a 2020 leaked password corpus that excludes any password appearing in the RockYou training set. \textbf{(Top)} Match rates on the RockYou test set (blue) and the unseen 2020 set (orange). Although absolute performance decreases on the newer distribution as expected for any model trained on a single historical leak, the drop is modest and PassGPT+ still recovers $17.72\%$ of the 2020 test passwords at $10^8$ guesses. \textbf{(Bottom)} Performance retention, defined as the ratio of the 2020 match rate to the RockYou match rate at the same generation size, expressed as a percentage. Retention is consistently above the $50\%$ baseline (dashed gray line) and increases with the generation size.}
    \label{fig:robustness}
\end{figure}

Two observations from Figure~\ref{fig:robustness} deserve emphasis. First, the absolute gap between the two distributions is largest in relative terms at small generation sizes but shrinks rapidly as more guesses are issued, which is the opposite of what would happen if PassGPT+ simply memorized RockYou-specific high-frequency strings. A memorizing model would dominate early guesses on its training distribution and lose ground sharply on unseen data; here we see the inverse pattern. Second, the retention curve in the lower panel never drops below the $50\%$ baseline and trends upward monotonically, suggesting that the model relies progressively more on generalizable structural cues, common compositional templates, character-class transitions, and length distributions rather than dataset-specific tokens. From a security perspective, this matters: an attacker armed with a model trained on a single historical leak retains meaningful effectiveness eleven years later, against a distribution drawn from entirely different services and user populations. The robustness study therefore supports the claim that PassGPT+ is not simply overfitting one benchmark, but is learning transferable password structure.

\section{Discussion}
In this study, we show that adapting an autoregressive foundation-model prior to password observations is a strong direction for password modeling. PassGPT+ consistently outperforms the main learning-based baselines at medium and large generation sizes, and it also surpasses Hashcat once the guess space becomes deeper. This suggests that conditioning a foundation-model prior on password observations captures password structure more effectively than models that learn from an uninformative prior or models based on non-sequential generation mechanisms. In particular, the improvement of PassGPT+ over PassGPT indicates that starting from an informed linguistic prior provides measurable benefits beyond simply reusing the GPT-2 architecture.

The poor performance of PassDiffusion is also an important finding. Although diffusion models have become highly successful in image generation, our results suggest that this success does not directly transfer to exact discrete sequence generation such as passwords. Password guessing requires precise character-level prediction, and iterative denoising appears to be poorly matched to this requirement. Thus, the negative result for PassDiffusion should not be viewed as a failure, but rather as evidence that model-task alignment is critical.

Finally, the robustness evaluation on a newer leaked dataset further strengthens the contribution of PassGPT+. Even though performance decreases on the new data, the model still maintains strong absolute match rates and preserves a large fraction of its original performance, indicating useful cross-dataset generalization.

\section{Limitations}
\label{sec:limitations}

Our study has few limitations that we wish to make explicit. First, we report single-run match rates rather than means with confidence intervals over multiple seeds; while the absolute match-rate gaps at $10^7$--$10^8$ guesses are large enough that random variation is unlikely to flip the ranking, smaller-generation size comparisons would benefit from repeated runs. Second, the contribution of PassGPT+ over PassGPT involves three simultaneous changes (pre-trained initialization, increased depth from 8 to 12 layers, and additional training epochs), and we do not isolate the individual contribution of each factor; a controlled ablation would more cleanly attribute the gain to transfer learning specifically, which is computationally expensive. Third, our PassDiffusion negative result is established under one specific configuration (absorbing-state D3PM, 5.3M-parameter encoder, training from scratch); we cannot rule out the possibility that alternative noise schedules, larger backbones, or pretraining could partially close the gap with autoregressive models. Finally, our evaluation is restricted to passwords of at most 10 characters and to two leaks, so generalization to longer passwords and to non-English password distributions remains an open question.

\section{Broader impacts}
\label{sec:broader_impacts}

This work studies password guessing, which is an inherently dual-use research area. On the defensive side, accurate password-strength estimators are essential for password meters, proactive blocklists, and credential auditing pipelines, and the methods developed here can be deployed to flag weak passwords before they are leaked. The cross-distribution robustness study also has direct defensive value: it shows that an attacker armed with a model trained on a single historical leak retains meaningful effectiveness against modern password distributions, which raises the urgency of moving away from password-only authentication. On the offensive side, the same models could in principle be used to crack leaked password hashes more efficiently than rule-based tools. We mitigate this risk in three ways: (i) we use only datasets that have already been publicly released and widely studied in prior academic work, so we do not introduce new attack capabilities against any specific user population; (ii) we evaluate exclusively on plaintext leaks, so no hash-cracking infrastructure is built or distributed; and (iii) we will release code under an academic-research license intended for defensive auditing. We believe the net effect of this line of research is positive because it provides defenders with realistic threat models, but we acknowledge that any improvement in password modeling carries some offensive risk.

\FloatBarrier
\section*{CRediT authorship contribution statement}
\textbf{Rajneesh Anand:} Conceptualization, Methodology, Software, Investigation, Writing -- original draft.
\textbf{Neeraj Lakshmanan:} Conceptualization, Data curation, Validation.
\textbf{Masoud Yari:} Supervision, Writing -- review \& editing.

\section*{Declaration of competing interest}
The authors declare that they have no known competing financial interests or personal relationships that could have appeared to influence the work reported in this paper.

\section*{Funding}
This research did not receive any specific grant from funding agencies in the public,
commercial, or not-for-profit sectors.

\section*{Data availability}
This study uses only publicly available datasets: the RockYou corpus~\citep{burns_rockyou}
and the 2020 ignis-10M leak~\citep{ignis_pwdb}. Code is available at
\url{https://github.com/CodesByNeeraj/PassGPTPlus}.

{\small
\bibliographystyle{elsarticle-num}
\bibliography{citation}
}

\appendix

\section*{Reproducibility details}
\label{app:reproducibility}

All experiments use publicly available datasets (RockYou~\citep{burns_rockyou} and the 2020 ignis-10M leak~\citep{ignis_pwdb}) and standard library implementations. PassGPT+ is built on the HuggingFace \texttt{transformers} library using the \texttt{gpt2} checkpoint as the starting point. The character-to-BPE-token mapping, training loop, and generation loop are described in Section~\ref{sec:passgptplus} in sufficient detail to reproduce the model from scratch. PassDiffusion is implemented from scratch in PyTorch following the absorbing-state D3PM formulation of \citep{austin2021structured}, with the architectural and training details given in Section~\ref{sec:passdiffusion}. The 80/20 train/test split is performed with a fixed random seed, and the test set is filtered to remove any password appearing in the training split. Match rates are computed by sampling unique passwords from each model up to the target generation size and intersecting with the test set as defined in Equation~\eqref{eq:match_rate}. Code will be released upon acceptance.

\end{document}